\documentclass[prd,amsbook,showpacs,nofootinbib,tightenlines,preprintnumbers,%
superscriptaddress,twocolumn]{revtex4}
\usepackage{epsfig,graphics}
\usepackage{amssymb,amsmath,amsthm,graphicx,psfrag}
\graphicspath{{./Figures/}}

\newcommand{\beqn}{\begin{eqnarray}}
\newcommand{\eeqn}{\end{eqnarray}}
\newcommand{\be}{\begin{equation}}
\newcommand{\ee}{\end{equation}}

\begin{document}
\preprint{HU-EP-08/35, ITEP-LAT/2008-17}
\title{The dyonic picture of topological objects in the deconfined phase}
\author{V.~G.~Bornyakov}
\affiliation{Institute for High Energy Physics,
Protvino, 142281, Russia}
\affiliation{Institute for Theoretical and Experimental Physics,
B. Cheremushkinskaya 25, Moscow 117259, Russia}
\author{E.-M. Ilgenfritz}
\affiliation{Institut f\"ur Physik, Humboldt-Universit\"at zu Berlin,
Newtonstr. 15, D-12489 Berlin, Germany}
\affiliation{Institut f\"ur Physik, Karl-Franzens-Universit\"at Graz, 
Universit\"atsplatz 5, A-8010 Graz, Austria}
\author{B. V. Martemyanov}
\affiliation{Institute for Theoretical and Experimental Physics,
B. Cheremushkinskaya 25, Moscow 117259, Russia}
\author{M. M\"uller-Preussker}
\affiliation{Institut f\"ur Physik, Humboldt-Universit\"at zu Berlin,
Newtonstr. 15, D-12489 Berlin, Germany}

\vspace{1.0cm}
\date{\today}

\begin{abstract}
In the deconfinement phase of quenched $SU(2)$ Yang-Mills theory the
spectrum and localization properties of the eigenmodes of the overlap 
Dirac operator with antiperiodic boundary conditions are strongly 
dependent on the sign of the average Polyakov loop, $\langle L \rangle$. 
For $\langle L \rangle > 0$ a gap appears with only few, highly localized 
topological zero and near-zero modes separated from the rest of the spectrum. 
Instead of a gap, for $\langle L \rangle < 0$ a high spectral density of 
relatively delocalized near-zero modes is observed. In an ensemble of 
positive $\langle L \rangle$, the same difference of the spectrum appears 
under a change of fermionic boundary conditions. 
We argue that this effect and other properties of near-zero modes can be 
explained through the asymmetric properties and the different abundance of 
dyons and antidyons -- topological objects also known to appear, however 
in a symmetric form, in the confinement phase at $T < T_c$ as constituents 
of calorons with maximally nontrivial holonomy.
\end{abstract}


\maketitle


\section{Introduction}
\label{sec:introduction}
\vspace{-0.2cm}

The study of topological objects at nonzero temperature on the basis of 
smeared $SU(2)$ lattice fields~\cite{IMMV06} has suggested the following 
picture of the topological content of $SU(2)$ lattice gauge theory. 
At low temperatures topological objects are represented by nondissociated 
calorons with maximally nontrivial holonomy~\cite{KvB-1,KvB-2,LL}.
(see also Ref.~\cite{Brower}).
With increasing temperature their composite nature becomes recognizable. 
They start to dissociate into dyons of topological charge $\pm 1/2$. 
Approaching the critical temperature $T_c$ (of the deconfining phase 
transition) from below, approximately 50 percent of the calorons become
dissociated, retaining their symmetric properties.
Above the critical temperature, a nonzero expectation value of the averaged 
Polyakov loop $\langle L \rangle \equiv \langle \overline{P(\vec{x})} \rangle$,
where $\overline{P(\vec{x})}$ means the 3-space average of the local values 
$P(\vec{x})$,
is realized, apart from tunnelings changing the sign of $\overline{P}$ which
are suppressed in large volumes. In the result there appears 
an asymmetry of dyons with peaked values of the local Polyakov loop 
$P(\vec{x})$ differing in sign: light dyons with the local Polyakov loop
of same sign as $\langle L \rangle$ become the most abundant topological 
objects, while heavy dyons (and even more nondissociated calorons) are 
suppressed.

For the confinement phase, Diakonov and Petrov~\cite{Diakonov:2007nv} 
have developed a confining dyon gas picture. Although it is a model either 
for purely selfdual (or antiselfdual) constituents, it explains all
essential features of that phase. For the numerical success it is important
that both self\-dual and antiselfdual gases cooperate without interaction.
This attractive model is, however, difficult to substantiate in lattice 
simulations because at low temperature the distances between caloron 
constituents are hard to be resolved.

Whether dissociated~\cite{Diakonov:2007nv} or undissociated bound in 
calorons~\cite{Gerhold}, dyons and antidyons in the confinement phase with 
$\langle L \rangle = 0$ are completely symmetric concerning the abundance 
between all four sorts of constituents: selfdual dyons $M$, $L$ 
(with positive topological charge, i.e. equal-sign electric and magnetic 
charge) forming calorons and antiselfdual antidyons $\overline{M}$, 
$\overline{L}$ (with negative topological charge, i.e. opposite-sign electric 
and magnetic charge) forming anticalorons. This nomenclature was coined by
Diakonov and Petrov~\cite{DP2003} in a paper, where they considered the role 
of these Bogomol'nyi-Prasad-Sommerfield~\cite{BPS} dyons not in 
nonsupersymmetric Yang-Mills theory but in ${\cal N}=1$ supersymmeric 
$SU(2)$ Yang-Mills theory.

What do we actually know about single caloron solutions ?
The semiclassical amplitude for $SU(2)$ calorons has been calculated 
in Ref.~\cite{DGPS2004}. There it has been shown that they are stable 
above $T_c \approx \Lambda_{\rm QCD}$ for holonomy 
$|\langle L \rangle| > 0.787597$. This actually leaves the room open 
for a dyon gas model as sketched above to describe the confined phase, 
but wasn't meant as an argument that the deconfined phase would be 
correctly described as a gas of ``atomic'' calorons and anticalorons. 

A naive picture, assuming that {\it undissociated calorons} dominate the 
{\it deconfined phase}, would associate the {\it onset of confinement} 
with the {\it dissociation of calorons}. The high-temperature phase would 
be a gas of calorons in their undissociated form, with a radius decreasing
with rising $T$ and adapted in shape to the respective $\langle L \rangle$. 
This point of view was taken in Ref.~\cite{Gerhold} as far as the deconfined 
phase was considered and found to disagree with the lattice observations. 
It ignores the possibility that the emerging nonzero value of the average 
Polyakov loop generates an enormous asymmetry between (anti-) dyons with a 
local Polyakov loop $P({\vec x})$ (in the center of their action or topological 
charge lumps) having different sign. If they were not necessarily bound in 
a caloron, ``light dyons'' with a peak value of the Polyakov loop 
$P({\vec x})$ equal in sign to $\langle L \rangle$ 
could become the most abundant objects carrying only a small amount of 
topological charge, while 
``heavy dyons'' with a local Polyakov loop $P({\vec x})$ opposite in sign 
to $\langle L \rangle$ 
could be heavily suppressed 
while carrying a relatively 
large topological charge close to $|Q|=1.$ Indeed, the higher action 
$S_{\rm heavy} = \frac{16\pi^2}{g^2}\left(\frac{1}{2}-\omega\right)$ (if
$\langle L \rangle= \cos\left(2 \pi \omega\right) > 0$) of the heavy dyons 
would be a natural explanation of their statistical suppression compared to 
light dyons with their lower action $S_{\rm light} = \frac{16\pi^2}{g^2} \omega$ 
(see section \ref{sec:spectra} for definition of $\omega$). 

The analysis of topological objects by means of overlap fermions has confirmed 
the observations made for smeared lattice fields at the temperature of the 
thermal phase transition~\cite{BIMMMV07}. Since that time, an extended and 
model-independent investigation of overlap fermion spectra for $SU(2)$ below 
and above $T_c$ (up to $T = 2~T_c$) was performed~\cite{BLMPIM08}. 
The fermionic eigenmodes (and their spectral density) show some striking 
peculiarities above $T_c$ that our dyonic picture of the topological content 
of $SU(2)$ lattice gauge theory seems to be able to explain.
In the time since our paper~\cite{BIMMMV07} was written, our own analysis 
of topological objects by means of overlap fermions has progressed and 
concentrated on the high-temperature phase, not only on properties of 
individual overlap modes, but also on the possibility to 
extract~\cite{Weinberg} topological properties of the gauge field. 
In this paper we present the analysis of topological clusters above $T_c$ 
which again requires to use the UV filtered definition of the topological 
density. In this respect, the present work is a direct continuation of 
the previous one on calorons and dyons at the thermal phase 
transition~\cite{BIMMMV07}. 
The knowledge of the spectral density and localization properties of 
individual modes, now detailed for both signs of 
$\langle L \rangle$ for fixed, antiperiodic temporal boundary 
conditions~\cite{BLMPIM08}
or -- vice versa -- depending on the temporal boundary conditions for 
fixed $\langle L \rangle > 0$ 
as considered in this paper~\footnote{It is clear that for the Dirac operator 
spectrum changing sign of the averaged Polyakov loop is equivalent to changing 
boundary conditions from antiperiodic boundary conditions to periodic boundary 
conditions.}, corroborates the interpretation of the present cluster results 
that has emerged in the meantime.

In our previous work~\cite{BIMMMV07} concentrating on topological clusters we 
considered $SU(2)$ gluodynamics on $20^3\times 6$ lattice using the tree-level 
improved Symanzik action at $\beta_{\rm imp}=3.25$ corresponding to the thermal 
phase transition. Here we use a lattice of size $20^3\times 4$ at the same 
$\beta_{\rm imp}$, i.e. we are studying gluodynamics at a temperature 
$T = 1.5~T_c$. Our analysis is based on 67 equilibrium configurations.
For each configuration we have obtained the 20 lowest eigenmodes of the
Dirac operator.

The paper is organized as follows: in Sect.~\ref{sec:spectra} we analyze the 
properties of overlap fermion spectra above $T_c$ as found in this paper and 
in Ref.~\cite{BLMPIM08}
on the basis of the dyonic picture of the topological content of $SU(2)$ 
lattice gauge theory. In Sect.~\ref{sec:localization} we discuss the reason 
for the enormous difference in localization~\cite{BLMPIM08} of the eigenmodes 
in the gap region between the two types of boundary conditions or between the 
two signs of the average Polyakov loop, respectively. In Sect.~\ref{sec:clusters} 
we investigate topological clusters and their respective degree of 
(anti-)selfduality following the idea of Ref.~\cite{Gattringer} by constructing 
the topological charge density and the field strength tensor~\cite{Weinberg} 
in an UV filtered way from the lowest fermion eigenmodes. 
In Sect.~\ref{sec:zeromodeprofile} additional dyonic signatures of topological 
objects are presented, as those connected to the profiles of fermion modes, 
those connected to the monopole content of topological clusters and those 
related to the profiles of the Polyakov loop throughout the clusters.
Finally, we conclude in Sect.~\ref{sec:conclusions}.

\section{Overlap fermion spectra}
\label{sec:spectra}
\vspace{-0.2cm}

The classical caloron solution with nontrivial holonomy consists of two dyons 
with oppositely peaked values of the local Polyakov loop $P({\vec x})$. 
This local field assumes a value equal to $1$ or $-1$~\cite{KvB-2} inside the 
constituents. The coincidence of the eigenvalues of the holonomy matrix is the
definition of a monopole. 
For the caloron solution the overlap Dirac operator with periodic boundary 
conditions has a zero mode localized on the dyon with positive central Polyakov 
loop while the operator with antiperiodic boundary conditions has a zero mode 
localized on the dyon with negative central Polyakov loop~\cite{ferm1,ferm2}.
Whether the zero mode jumps from one dyon to another when the boundary 
conditions are changed~\cite{GattringerSchaefer} depends only on the degree 
of separation between the constituents (dissociation). Even within a 
nonseparated lump of action the zero mode is able to oscillate inside the 
lump under a change of boundary conditions.

\begin{figure*}[!htb]
\begin{center}
\includegraphics[width=.45\textwidth]{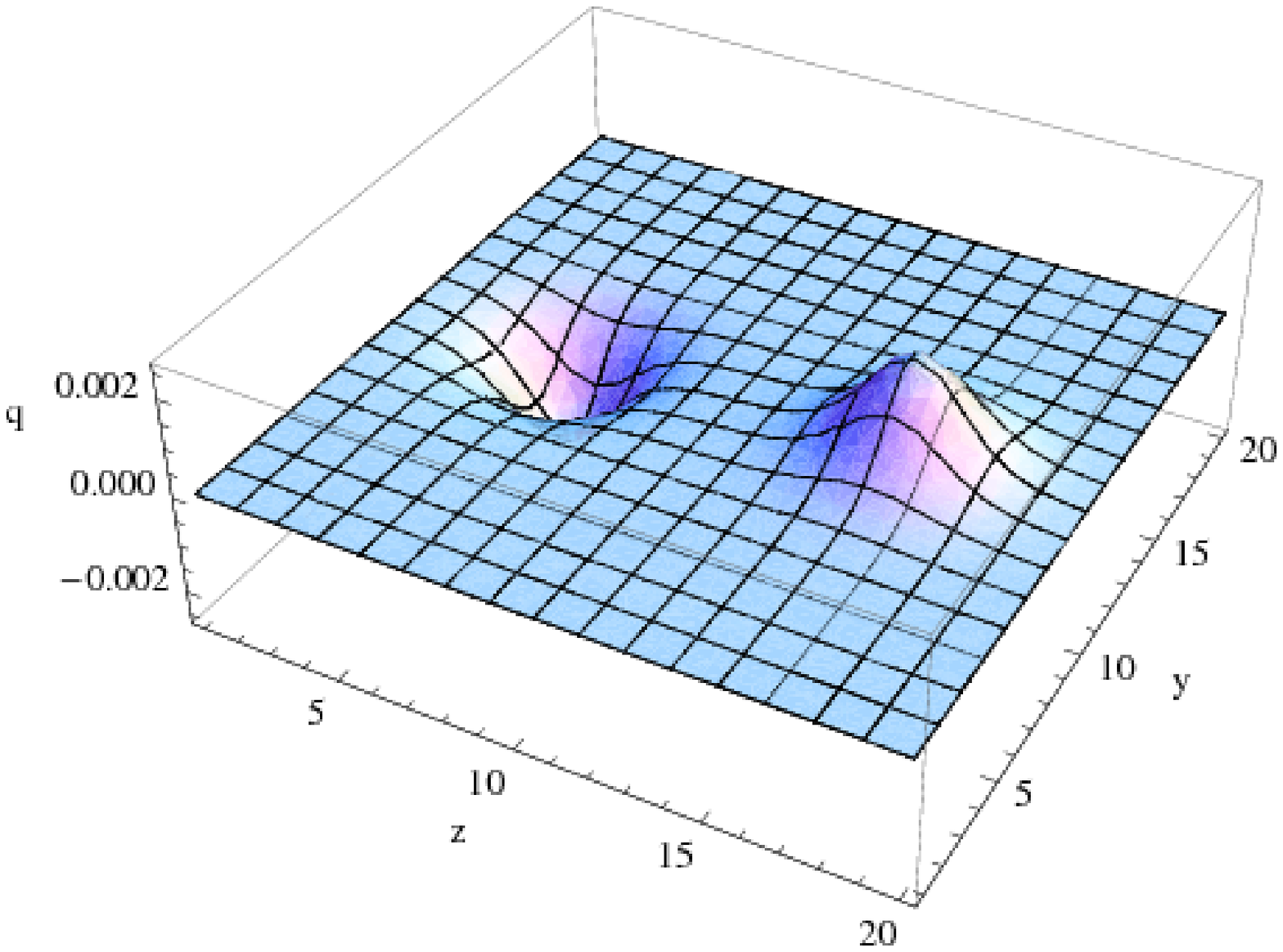}%
\hspace{0.5 cm}
\includegraphics[width=.45\textwidth]{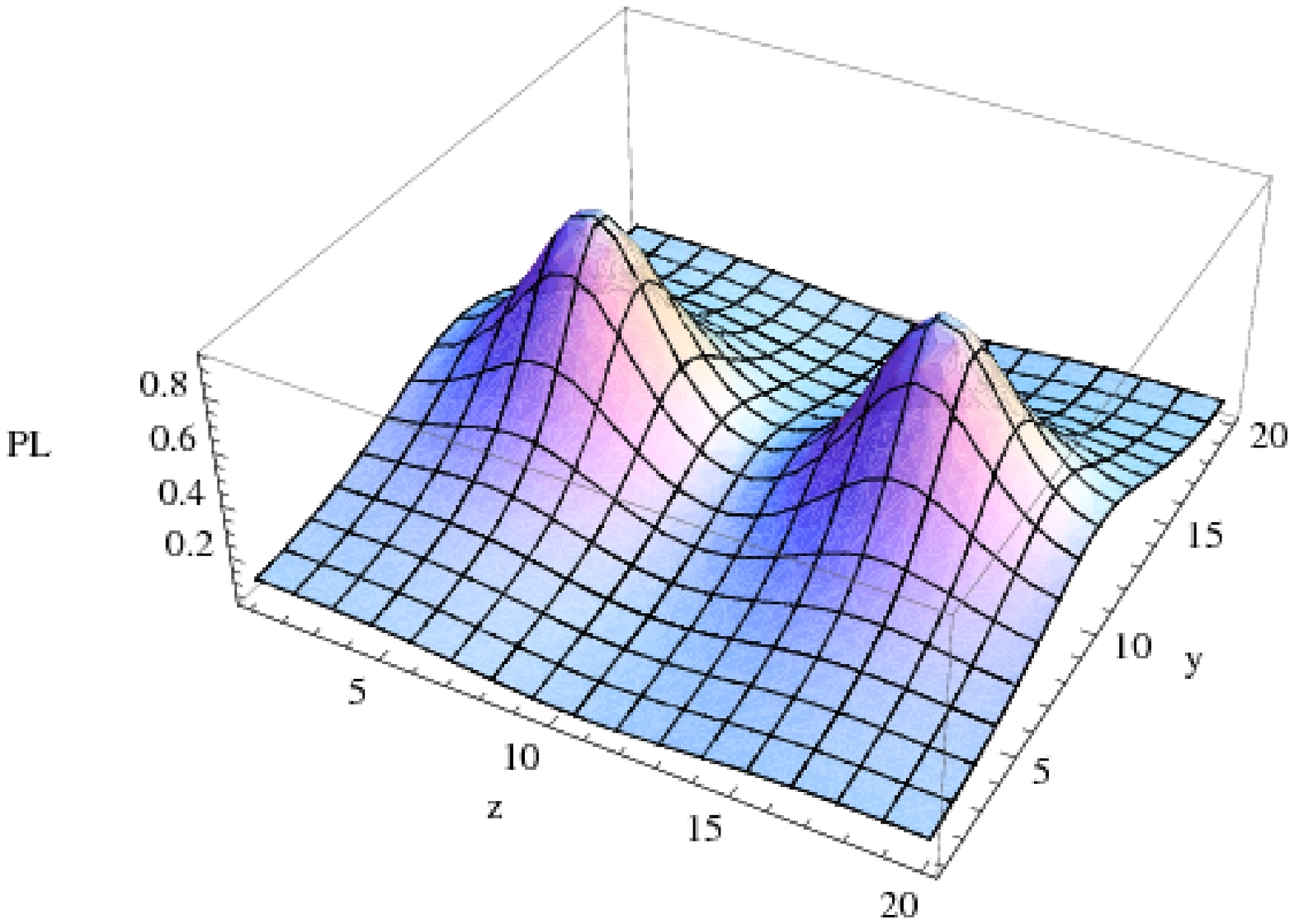}\\
(a)\hspace{3.5 cm}(b)\\
\vspace{2 cm}
\includegraphics[width=.45\textwidth]{dadspectra.eps}\\
(c)\\
\vspace{1 cm}
\includegraphics[width=.45\textwidth]{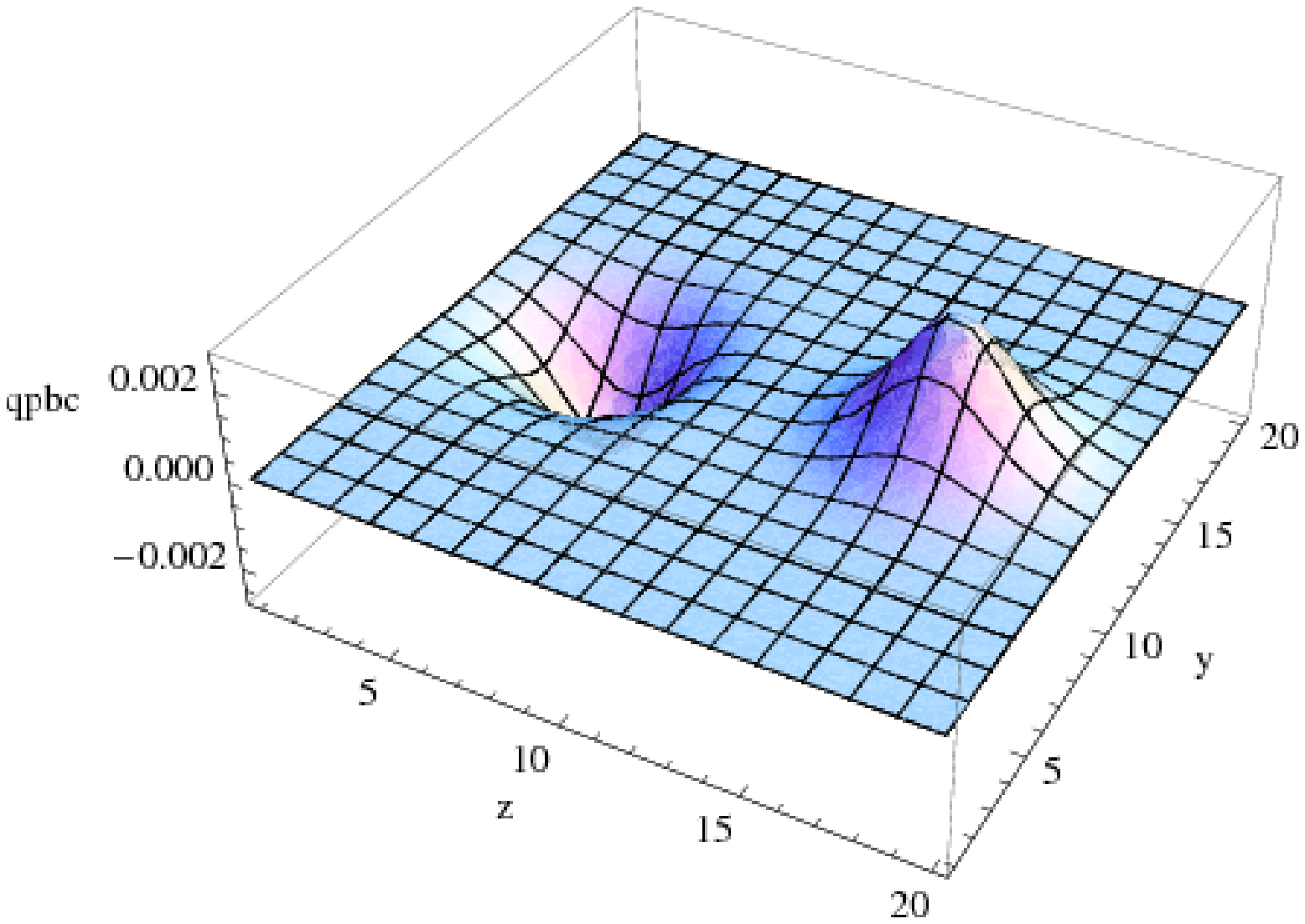}%
\hspace{0.5 cm}
\includegraphics[width=.45\textwidth]{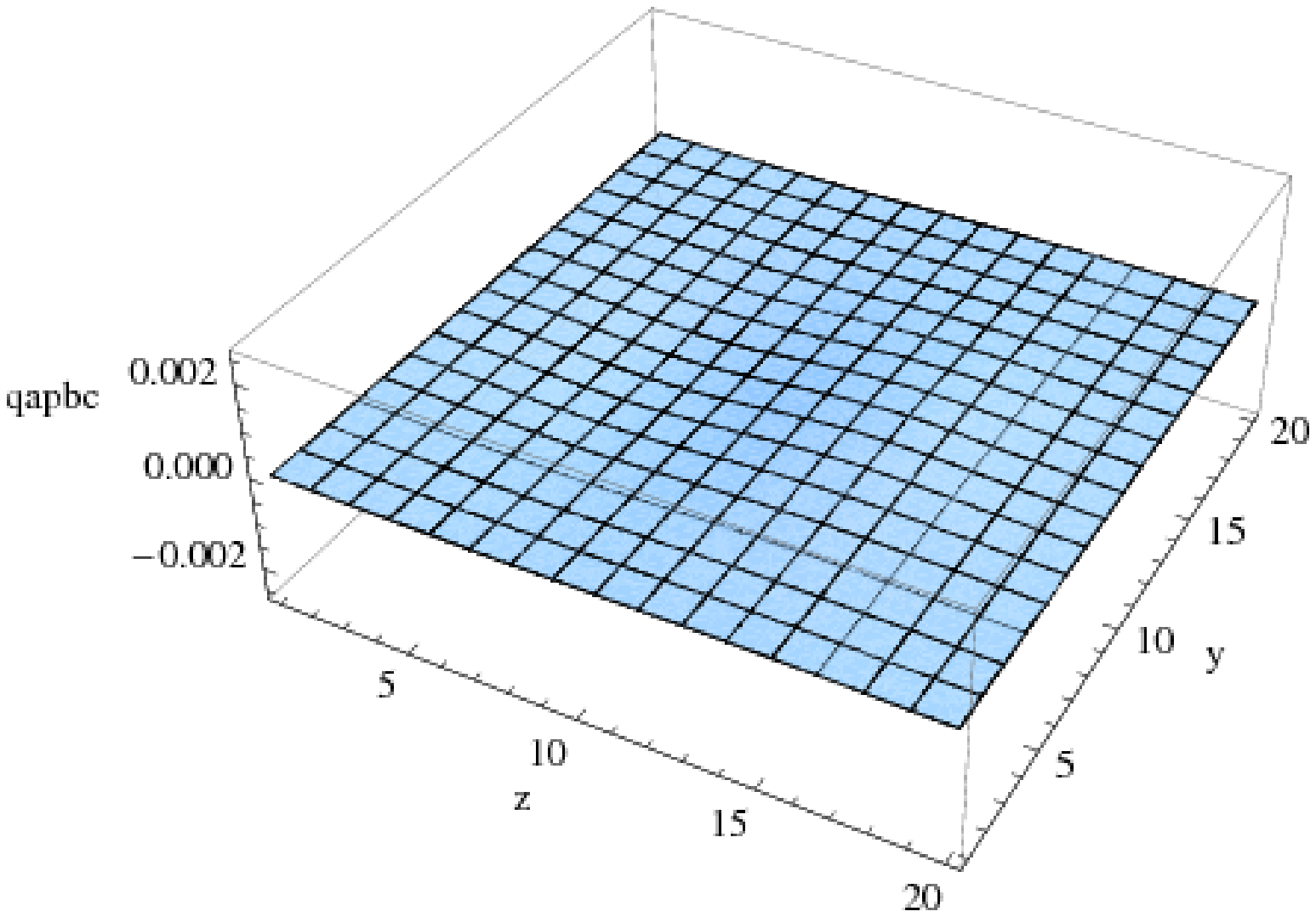}\\
(d)\hspace{3.5 cm}(e)\\
\end{center}
\vspace{-0.5cm}
\caption {(a) The gluonic topological charge density and (b) the 
local Polyakov loop for an artificially constructed dyon-antidyon pair. 
The spectrum (c) for overlap fermions with periodic (left) and 
antiperiodic (right) boundary conditions (restricted to 20 modes). 
Inside the circle around the origin the plot is 10 times magnified.
The fermionic topological charge density from these 20 lowest modes is 
shown for periodic (d) and antiperiodic (e) boundary conditions. 
The latter choice is blind for these constituents.}
\label{fig:dad}
\end{figure*}

A pair of dyon and antidyon with a same sign of the central Polyakov loop 
values $P({\vec x}_1)$ and $P({\vec x}_2)$ cannot constitute a classical 
solution. For field configuration having only topological objects of this 
type the Dirac operator has no exact zero mode because the total topological 
charge is zero. 
If both $P({\vec x}_i)>0$ or both $<0$, the Dirac operator with periodic 
(antiperiodic) boundary conditions has two near-zero modes 
which tend to become zero modes (of opposite chirality) only in the limit of 
infinite dyon-antidyon separation. This can be seen in Fig.~\ref{fig:dad}, 
where the spectrum 
for an artificially constructed dyon-antidyon pair is shown. A similar pair
has been obtained from generic lattice configurations by overimproved cooling 
in Ref.~\cite{InstantonQuarks} and shown in Fig. 11 therein.

\begin{figure*}[!htb]
\begin{center}\hspace*{-1.0cm}
\includegraphics[width=.65\textwidth]{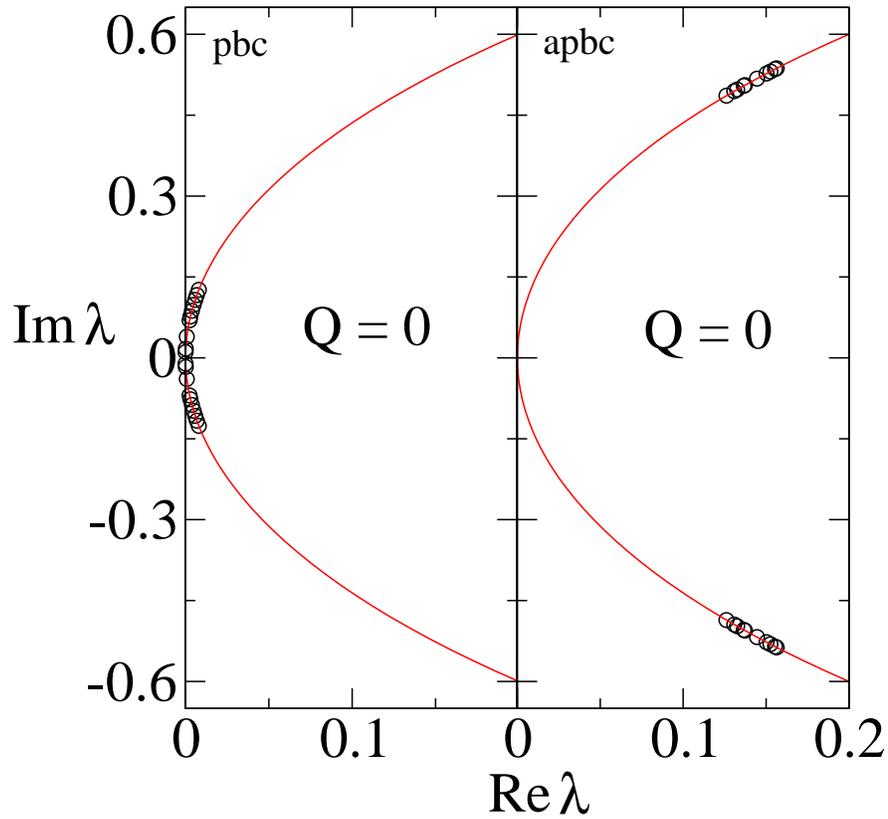}\\
\end{center}
\vspace{-0.5cm}
\caption {Periodic (left) and antiperiodic (right) overlap fermion spectra 
(including 20 modes) for one typical equilibrium Monte-Carlo configuration 
created in the deconfined phase with a positive average Polyakov loop 
$\langle L \rangle > 0$.}
\label{fig:spectra05}
\end{figure*}

Now let us compare the fermion spectrum for such an artificial dyon-antidyon
pair with those of equilibrium Monte-Carlo configuration in the deconfined 
phase of pure $SU(2)$ lattice gauge theory. This is shown 
in Fig.~\ref{fig:spectra05}
for periodic boundary conditions and antiperiodic boundary conditions.
The Monte Carlo configuration is a typical configuration from a sample with 
Polyakov loop $\langle L \rangle > 0$. It can be seen from 
Fig.~\ref{fig:spectra05} that the spectra for periodic boundary conditions and 
antiperiodic boundary conditions are very different. For periodic boundary 
conditions the spectrum has no gap while for antiperiodic boundary conditions 
there is a wide gap. Guided by the similarity between Fig.~\ref{fig:dad}c and 
Fig.~\ref{fig:spectra05} we can propose the following explanation of this 
difference. In the case of configurations with $\langle L \rangle > 0$  
light dyon-antidyon pairs with a Polyakov line peaking at $P({\vec x}) = +1$ 
appear in a large number and give rise to numerous near-zero modes in the 
spectrum of the Dirac operator with periodic boundary conditions. This eliminates 
the spectral gap completely, while in the case of the Dirac operator with 
antiperiodic boundary conditions only heavy dyon-antidyon pairs can produce 
near-zero modes, and such pairs are rare. 
This is reflected by the known fact of a gap opening for the case of 
antiperiodic boundary conditions. In fact, in equilibrium configurations 
there is a small number of (exceptional) near-zero modes seen. 
This has been first discovered in ~\cite{Edwards} and recently confirmed
in~\cite{BLMPIM08}. With increasing temperature, they become more and more 
separated from the rest of the spectrum (bulk) by the emerging gap and are 
decreasing in multiplicity~\cite{BLMPIM08}. 

\begin{figure*}[!b]   
\begin{center}\hspace*{-1.0cm}
\includegraphics[width=.65\textwidth]{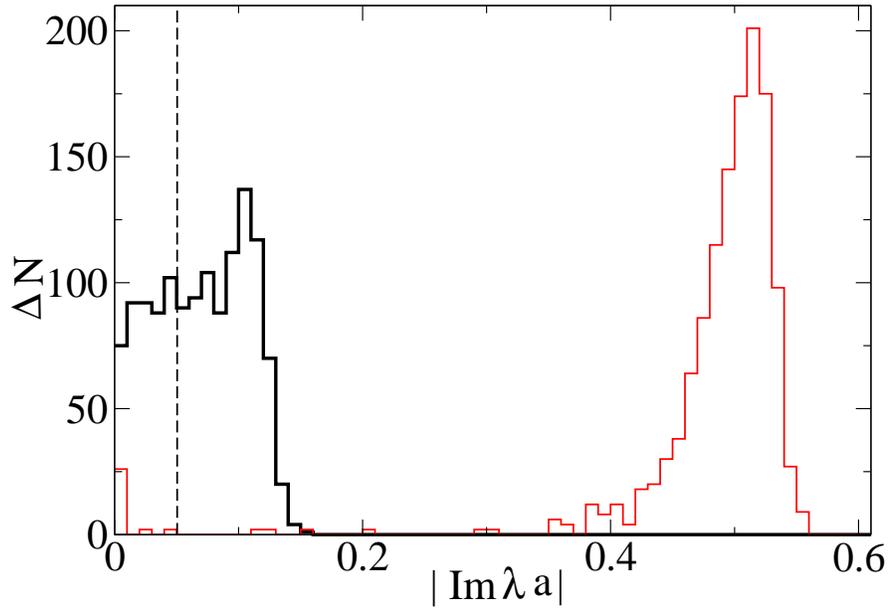}\\
\end{center}
\vspace{-0.5cm}
\caption{Overlap fermion spectra for periodic (black thick histogram) 
and for antiperiodic (red thin histogram) boundary conditions
obtained from 67 equilibrium 
Monte-Carlo configuration (all with positive averaged Polyakov loop)
in the deconfined phase. Shown are $(20\times 67 -54)$ periodic nonzero modes
and $(20\times 67 -52)$ antiperiodic nonzero modes.
The vertical dashed line shows the cut on near-zero modes, see the text for 
explanation.}
\label{spektra}
\end{figure*}

We identify the number of near-zero modes $n_{\rm nzm}$ ($N_{\rm nzm}$) found for
periodic (antiperiodic) boundary conditions (both with $\langle L \rangle >0$)
with the number of light (heavy) dyon-antidyon pairs.
The ratio of $n_{\rm nzm}/N_{\rm nzm}$ can be estimated from Fig.\ref{spektra}
where the spectra of 20 lowest eigenmodes are shown for both boundary conditions.
Choosing the cut on near-zero modes as 
$|{\rm Im}\lambda| < 0.05/a \approx 100 {\rm~MeV}$ shown
on the Fig.\ref{spektra} by the vertical dashed line we get 
$n_{\rm nzm}/N_{\rm nzm} \approx 15$. 
This number cannot be exact because in 
the case of periodic boundary conditions near-zero modes are not clearly
separated by a gap from the rest of the spectrum (the bulk). 
Therefore some of the modes counted in $n_{\rm nzm}$ might actually belong 
to the bulk. With this reservation in mind, one could conclude from the observed 
in Ref.~\cite{BLMPIM08} increasing with temperature spectral density of the periodic 
Dirac operator that the ratio $n_{\rm nzm}/N_{\rm nzm}$ is 
also increasing with temperature.

An alternative estimate for the number of light dyon-antidyon pairs, 
$n_{\rm nzm}$, can be made using the following observation. In a dilute gas of 
light and heavy dyons~\footnote{In the following ``dyon'' will be 
used without difference for carriers of positive and negative topological 
charge.} each object independently contributes to the square of the topological
charge $Q^2$ such that the sum over all configurations can be presented by
\be
\sum Q^2  = q^2  n + (1-|q|)^2  N \, ,
\label{eq:Q2}
\ee
where $n =n_{\rm zm}+n_{\rm nzm}$ is the total number of zero and near-zero 
modes for the Dirac operator with periodic boundary conditions, 
i.e. the total number of light dyons summed over all configurations, and 
$N =N_{\rm zm}+N_{\rm nzm}$ is the total number of zero and near-zero modes 
for the Dirac operator with antiperiodic boundary conditions, i.e. the total 
number of heavy dyons summed over all configurations. 
Thereby, $|q|$ is a fractional topological charge of a light dyon and the 
complement $1-|q|$ that of a heavy dyon. We remind that we are discussing 
an ensemble with positive averaged Polyakov loop $\langle L \rangle > 0$ 
which is actually the case for the configurations analyzed in the present paper.
For our set of 67 configurations we find that $\sum Q^2 = 78$, 
$N_{\rm zm} =52$, $N_{\rm nzm} =30$, and $n_{\rm zm} =54$.
We estimate the ratio between the topological charges of light and heavy 
dyons by the formula 
\be
|q|=2\omega
\ee
known from the analytical caloron solution:
\be
|q| : (1-|q|) = \omega : (1/2 -\omega) \, ,
\ee
where $\omega$ is the parameter of holonomy $H = \cos (2\pi\omega)$, and 
we identify the holonomy $H$ with the value of the average Polyakov loop 
$\langle L \rangle \approx  0.3$. We obtain $|q| \approx 0.4$ and 
$n_{\rm nzm} \approx 250$. Another estimate is based on the number
$|q| \approx 0.3$ that correspond to the maximum of the heavy dyon mass
distribution $1-|q| \approx 0.7$ (see Sect.\ref{sec:zeromodeprofile}).
Then $n_{\rm nzm} \approx 360$. These give a ratio 
$8.3 < n_{\rm nzm}/N_{\rm nzm}< 12  $ in acceptable agreement with the 
value $15$ obtained above comparing the number of low lying modes below 
$100 {\rm~MeV}$ (see Fig.\ref{spektra}), especially if one takes into account 
that we could have overestimated the number $n_{\rm nzm}$ of light nonzero 
modes in the case of periodic boundary conditions.

\section{Localization of eigenmodes}
\label{sec:localization}
\vspace{-0.2cm}

Next we turn to the issue of localization. This is a very natural question to 
ask in an investigation of the fermionic spectrum and eigenmodes. It turns out, 
that -- adopting the view in terms of topological charge clusters -- we can 
also contribute to the understanding of the strong dependence of the 
localization of zero and near-zero eigenmodes on the boundary conditions that
was discovered in~\cite{BLMPIM08}.

We find that for zero modes, the average inverse 
participation ratio (IPR) of antiperiodic zero modes is 
equal to $110$, while the IPR of periodic zero modes  is equal to $4.75$. 
For positive $\langle L \rangle $
-- as in our case -- the stronger localization of antiperiodic zero modes 
compared to periodic ones is easy to understand in the dyonic picture of 
the deconfined phase.

Each zero mode out of our $N_{\rm zm} =52$ antiperiodic zero modes is 
accompanied on average by $N_{\rm nzm}/(2 N_{\rm zm}) \approx 0.3$ 
{\it pairs of antiperiodic near-zero modes}, whereas the corresponding ratio
for periodic boundary conditions according to the actual number $n_{\rm zm}=54$  
of periodic zero modes requires to extract from (\ref{eq:Q2}) the number of
periodic near-zero modes. Assuming $|q| \approx 0.4$ the ratio is obtained as
$n_{\rm nzm}/(2 n_{\rm zm}) \approx 2.3$.
According to our dyonic picture we can say that an antiperiodic zero mode is
spread out on average over $1.3=1+0.3$ identical (i.e. with the same sign 
of topological charge) heavy dyons, while a periodic zero mode is spread out 
on average over $3.3=1+2.3$ identical light dyons. Moreover, taking into 
account that the radii of light and heavy dyons are different and follow, 
correspondingly to a caloron formula, the proportionality rule 
$r : R = (1-|q|) : |q| \approx 1.5$, we shall expect that an antiperiodic 
zero mode is $\left(\frac{3.3}{1.3}\right)\cdot(1.5)^3 \approx 9$
times more localized than a periodic zero mode. 
Repeating this estimate assuming $|q| \approx 0.3$ the above localization ratio
is replaced by $43$.
The actual ratio of the average IPR between antiperiodic zero modes and periodic 
zero modes for our ensemble, $110/4.75$, is halfway between the two estimates.

In Ref.~\cite{BLMPIM08} a tendency has been found 
that the localization of antiperiodic zero modes together with 
$\langle L \rangle > 0$ is increasing with increasing temperature, while the 
localization of antiperiodic zero modes in the presence of 
$\langle L \rangle < 0$ or, equivalently, periodic zero modes in the presence 
of $\langle L \rangle > 0$, is decreasing with increasing temperature. 
This qualitatively corresponds to the dyonic picture where the number ratio 
between heavy and light objects (dyons and antidyons) decreases and the size 
ratio between light and heavy objects increases with increasing temperature. 
This tendency is incorporated in the results obtained here by the temperature 
dependence of $\omega$ and $q$.

The properties of the above mentioned 
$M$, $L$, $\overline{M}$ and $\overline{L}$ dyons and antidyons are 
summarized with respect to the dependence on $\omega$ in 
Table~\ref{tab:dyontable}.
\begin{table}
  \begin{center}
    \begin{tabular}{|c|c|c|c|c|c|c|}
\hline
    type      & $P$  & $Q$   & $e$ & $m$ & action $\propto$ & size $\propto$ \\
\hline
    $M$       & $+1$ & $> 0$ & $\pm$ & $\pm$ & $\omega$         & $(1/2 - \omega)$   \\
    $L$       & $-1$ & $> 0$ & $\mp$ & $\mp$ & $(1/2 - \omega)$ & $\omega$           \\
    $\bar{M}$ & $+1$ & $< 0$ & $\pm$ & $\mp$ & $\omega$         & $(1/2 - \omega)$   \\
    $\bar{L}$ & $-1$ & $< 0$ & $\mp$ & $\pm$ & $(1/2 - \omega)$ & $\omega$           \\
\hline
\end{tabular}
\end{center}
\caption{Local Polyakov loop $P$, sign of topological charge $Q$, 
electric ($e$) and magnetic ($m$) charge of dyons and antidyons. 
Action and size are given for positive external Polyakov loop 
$\langle \overline{P(\vec{x})} \rangle = \cos\left(2 \pi \omega \right) > 0$.}
\label{tab:dyontable}
\end{table}

\section{Topological clusters }
\label{sec:clusters}
\vspace{-0.2cm}

Now let us turn to the properties of clusters with respect to the UV filtered
definition of the topological charge density. We remind the reader that we 
are considering here an ensemble of lattice configurations with average 
Polyakov loop $\langle L \rangle > 0$. The fermionic 
definition~\cite{Niedermayer} has an UV filtered 
variant~\cite{Horvath1,Horvath2}. Both have an {\it a priori} ambiguity with 
respect to the fermionic boundary conditions:
\be
q^{(b)}_{\lambda_{\rm cut}}(x) = - \sum_{|\lambda_{b}| \le \lambda_{\rm cut}} 
\left( 1 - \frac{\lambda_{b}}{2} \right) 
\psi^{(b)\dagger}_{\lambda_{b}}(x) \gamma_5 \psi^{(b)}_{\lambda_{b}}(x)  \, ,
\label{eq:truncated_density_bc}
\ee
with $b=p$ denoting periodic and $b=a$ denoting antiperiodic temporal boundary 
conditions. Although the total topological charge given by the number of zero
modes is not affected by the boundary condition as long as some smoothness 
properties of the gauge field are fulfilled~\cite{BIMMMV07}, we find that the
filtered density function really depends on the boundary condition $b$. 
This has allowed us in Ref.~\cite{BIMMMV07} to investigate the dyonic vs.
caloron structure for $T \lesssim T_c$ by measuring the amount of displacement 
of constituents with different sign of $P({\vec x})$.

For the high-temperature phase, we adopt one more difference in our procedure, 
depending on the type of boundary conditions. In the case of periodic boundary 
conditions we include all the 20 lowest modes which we computed. As discussed
above, in this case the near-zero modes cannot be strictly separated from the 
bulk of the spectrum. In the case of antiperiodic boundary conditions, 
however, we include only the zero and near-zero modes in the definition 
(\ref{eq:truncated_density_bc}) because they can be separated without any 
ambiguity from the bulk.

The antiperiodic boundary condition highlights the heavy constituents with 
negative local Polyakov loop, whereas the periodic boundary condition 
emphasizes the complementary light constituents with positive local Polyakov 
loop. In the same way as in Ref.~\cite{BIMMMV07} we now define clusters 
of topological charge as connected sets of lattice sites where the absolute 
value of topological charge density exceeds some cut, 
$|q(x)| > q_{\rm cut} = \frac{1}{5} \max_x(|q(x)|)$. The density
$q(x) = q^{(b)}_{\lambda_{\rm cut}}(x)$ is the adopted version of UV filtered
topological density defined by the respective selection of the 
modes~\footnote{The tail of the topological charge distribution with 
$|q(x)| < \frac{1}{5} \max_x(|q(x)|)$ is part of the volume where we expect 
the field strength tensor to deviate strongly from being selfdual or 
antiselfdual.}.

Furthermore, we have constructed from the fermionic modes the field strength 
tensor~\cite{LAH} 
\be
c^T~F^a_{\mu\nu}(x) =  
\sum_{|\lambda_{b}| \le \lambda_{\rm cut}} \lambda_{b} 
\psi^{(b)\dagger}_{\lambda_{b}}(x)~\sigma_{\mu\nu} \tau^a 
\psi^{(b)}_{\lambda_{b}}(x)  \, ,
\label{eq:fieldstrength}
\ee
where $c^T = 0.0883$~\cite{LAH1} has been calculated for the full spectrum.
For $F^a_{\mu\nu}$ on the left hand side of Eq. (\ref{eq:fieldstrength}) we 
have omitted the labels $b$ and $\lambda_{\rm cut}$ in order to avoid clumsy 
formulae in what follows.

\begin{figure*}[!htb]
\begin{center}
\includegraphics[width=.8\textwidth]{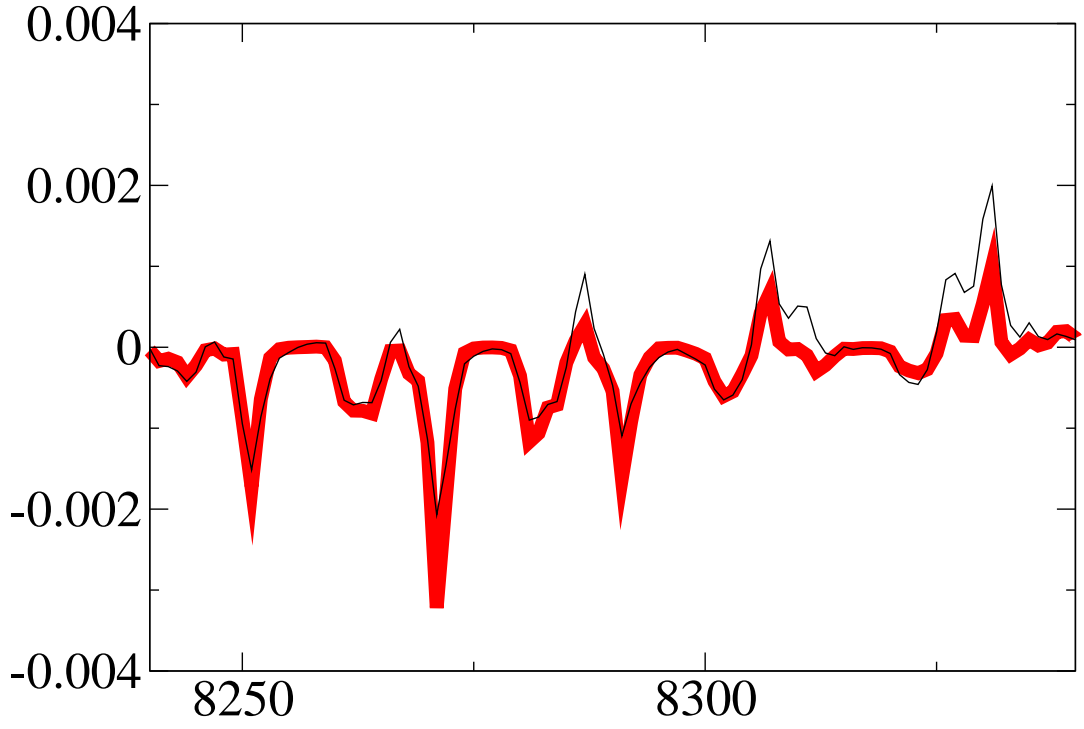}\\
\end{center}
\vspace{-0.5cm}
\caption {The topological charge density $q^{(p)}_{\lambda_{\rm cut}}(x)$ 
(thin line) and the rescaled topological charge density 
${q^F}^{(p)}_{\lambda_{\rm cut}}(x)$ (thick line) for subsequent lattice sites
along the $x$-axis in a typical lattice configuration.}
\label{fig:qfq}
\end{figure*}

The topological charge densities defined by (\ref{eq:truncated_density_bc}) on
one hand and by the field strength tensor on the other, through the scalar
product ${\vec E}^a \cdot {\vec B}^a$ of electric and magnetic field strength,
\be
q^F(x) = \frac{1}{32\pi^2} F^a_{\mu\nu}(x) \tilde{F}^a_{\mu\nu}(x) \, ,
\label{eq:fieldstrength1}
\ee
differ for truncated sums over modes like in our case by many orders of 
magnitude. However, if suitably rescaled, the density (\ref{eq:fieldstrength1})
closely follows the density (\ref{eq:truncated_density_bc}).
Fig.~\ref{fig:qfq} shows this for an on-axis sequence of lattice sites in one 
particular configuration studied with periodic boundary conditions and taking
all 20 modes into account. A rescaling factor $r$ has been found by minimizing the 
quantity
\be
\Delta^2 =  \overline {\left(q^{(b)}_{\lambda_{\rm cut}}(x)
 - r\cdot q^F(x) \right)^2 }  \,,
\label{eq:chi2}
\ee
where the bar means averaging over all lattice sites. We found that the 
parameter $r$ is of the order $O(10^8)$. For the particular configuration 
presented in Fig.~\ref{fig:qfq} we have found the deviation 
$\Delta \approx 0.04 \max_x |q^{(b)}_{\lambda_{\rm cut}}(x)|$, i.e. it amounts
to only a few percent of the maximal density.

\begin{figure*}[!htb]
\begin{center}
\includegraphics[width=.8\textwidth]{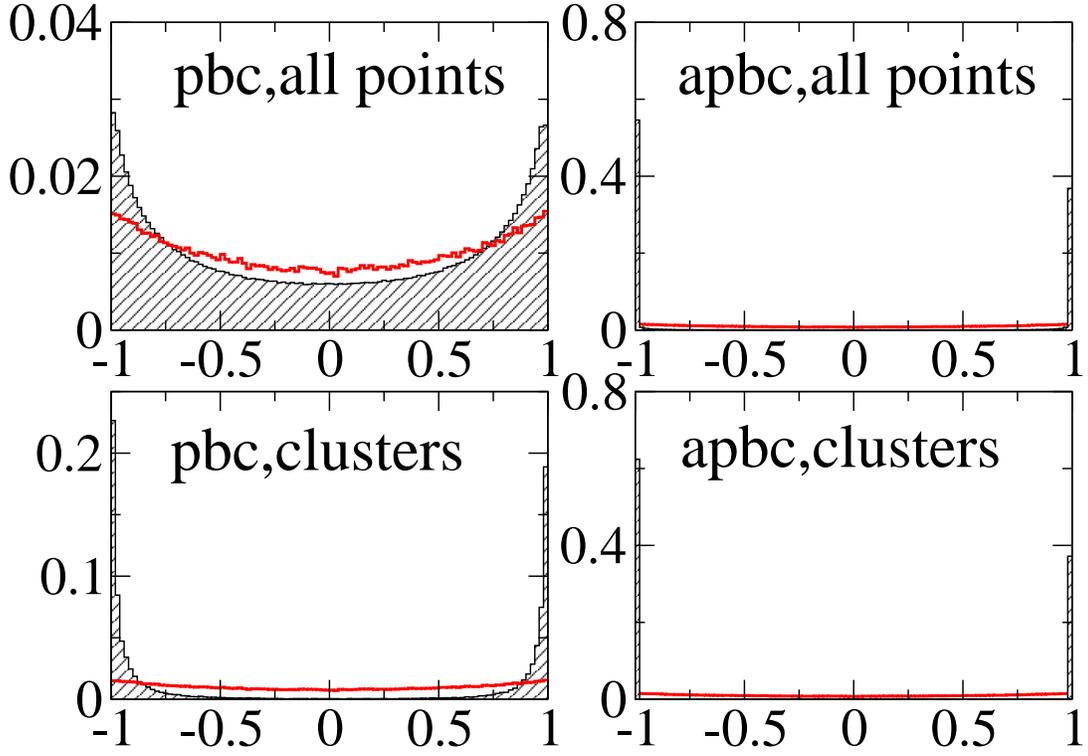}\\
\end{center}
\vspace{-0.5cm}
\caption {Histograms with respect to the (anti)selfduality $R$ 
(\ref{eq:R})
of all lattice sites (upper row) and within the interior of topological
charge clusters (bottom row). Left column: the 20 lowest modes with periodic 
boundary conditions have been used for the construction of the UV filtered 
topological density (\ref{eq:truncated_density_bc}) and the 
UV filtered field strength tensor (\ref{eq:fieldstrength}).
Right column: only zero and near-zero modes (below the gap) for antiperiodic 
boundary conditions have been used in constructing the UV filtered quantities.
In this case the reconstructed field strength in all sites is either selfdual 
or antiselfdual, whether they belong to topological clusters (with more than 
1/5 of the maximal density) or not. The fat line (red in the colored version) shows the histograms refering 
to a random assignment of $R$ (see text).}
\label{fig:sd}
\end{figure*}

In order to focus on the dyonic nature of the topological objects detected by 
the respective number of modes, we tested the degree of 
(anti)selfduality~\cite{Gattringer} of the topological clusters as mapped out 
by the topological density (\ref{eq:truncated_density_bc}). Site by site we 
considered the quantity
\be
R = \frac{4}{\pi}\arctan
\frac{F^a_{\mu\nu} \cdot F^a_{\mu\nu} - F^a_{\mu\nu} \cdot \tilde{F}^a_{\mu\nu}}
     {F^a_{\mu\nu} \cdot F^a_{\mu\nu} + F^a_{\mu\nu} \cdot \tilde{F}^a_{\mu\nu}}
 -1 \, .
\label{eq:R}
\ee
Analogously to studies of the local chirality of fermionic modes, this quantity 
equals to $-1$ ($+1$) for a strictly selfdual (antiselfdual) field strength
tensor. 
The distribution of this quantity for {\it all lattice sites} on one hand and 
restricted to the {\it interior of the topological charge clusters} on the 
other is shown in Fig.~\ref{fig:sd}. The case of periodic boundary conditions, 
where the near-zero modes cannot be clearly separated from the background, is 
presented in the left panels of Fig.~\ref{fig:sd}: 
the interior of the topological 
charge clusters (shown left below) is indeed preferentially selfdual or 
antiselfdual. In contrast, taking all lattice sites into account (shown left 
above), the distribution with respect to the degree of (anti)selfduality can 
hardly be distinguished from that obtained for a random assignment. This
random reference case is represented by the fat line (red in the colored 
version) 
in the left panels of Fig.~\ref{fig:sd}. The random construction consists in 
replacing $F^a_{\mu\nu} \cdot \tilde{F}^a_{\mu\nu}$ by a random number $q$ 
sampled from the interval $[- s, + s]$ with 
$s = F^a_{\mu\nu} \cdot F^a_{\mu\nu}$.

In the case of antiperiodic boundary conditions we are able to unambiguously 
restrict ourselves to the zero and near-zero modes in the construction of
(\ref{eq:truncated_density_bc}) and (\ref{eq:fieldstrength}).
From the right panels of Fig.~\ref{fig:sd} it becomes obvious now why we 
associate these modes to the heavy dyons (again irrespective of the sign of the
topological charge density). Their contributions to 
(\ref{eq:truncated_density_bc}) and (\ref{eq:fieldstrength}) form what we can 
call ``selfdual and antiselfdual heavy dyons''. If the construction is limited 
to these modes the obtained field strength is constrained to be either selfdual
or antiselfdual, irrespective whether we consider all lattice sites 
(shown right above) or only the interior of the clusters of sufficiently 
large topological charge density compared to the maximal density (shown right
below). This is expressed by the fact that the right
panels of Fig.~\ref{fig:sd} both collapse to a sum of $\delta$-functions at 
$R=-1$ and $R=+1$. The height of the peaks reflects the ratio of positive and 
negative topological charges summed over the ensemble. For this case all 
$N_{\rm zm}+N_{\rm nzm}=52+30=82$ heavy zero and near-zero modes are in 
one-to-one correspondence to the observed clusters of topological charge. 

\section{Profiles of near-zero and zero modes}
\label{sec:zeromodeprofile}
\vspace{-0.2cm}

Last but not least we consider more in detail the shapes of the $82$ 
heavy zero and nonzero fermionic 
modes in comparison to analytical results available for dyons~\cite{ferm1}. 
Analytically, from the caloron solution, the profile of a zero mode localized 
on one of the constituent dyons is known to depend on two parameters of the
caloron: the holonomy parameter $\omega$ that defines the ``mass'' fraction 
of the constituent (also the fractional topological charge) as 
$m_{h} = 1-2\omega$ 
and the distance $d$ from the center of the accompanying complementary 
constituent with a ``mass'' fraction  $m_{l} = 2\omega$. Both parameters 
influence the nonstaticity of zero mode. The parameters $\omega$ and $d$ were 
found for every lattice zero and near-zero mode using the following procedure. 

\begin{figure*}[!htb]
\begin{center}
\includegraphics[width=.7\textwidth]{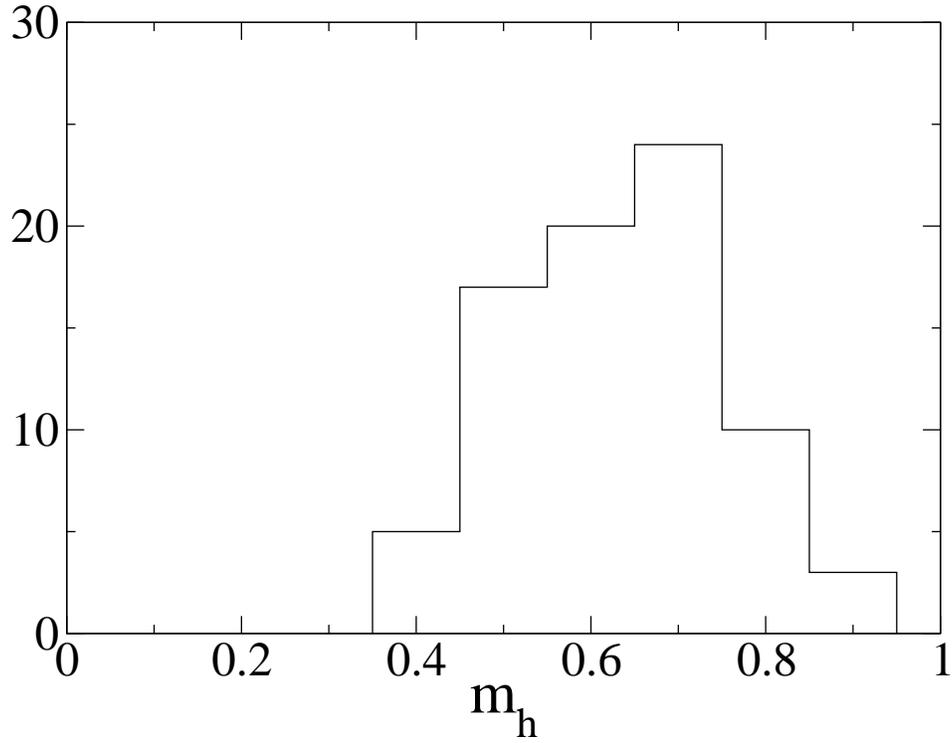}\\
\end{center}
\vspace{-0.5cm}
\caption {The distribution of the fractional action (modulus of topological 
charge) concentrated in heavy dyons, in our simulation represented by the 
$82$ zero and near-zero modes for antiperiodic boundary conditions.
Our ensemble has positive average Polyakov loop $\langle L \rangle > 0$.}
\label{fig:nudistr}
\end{figure*}

\begin{figure*}[!htb]
\begin{center}
\includegraphics[width=.8\textwidth]{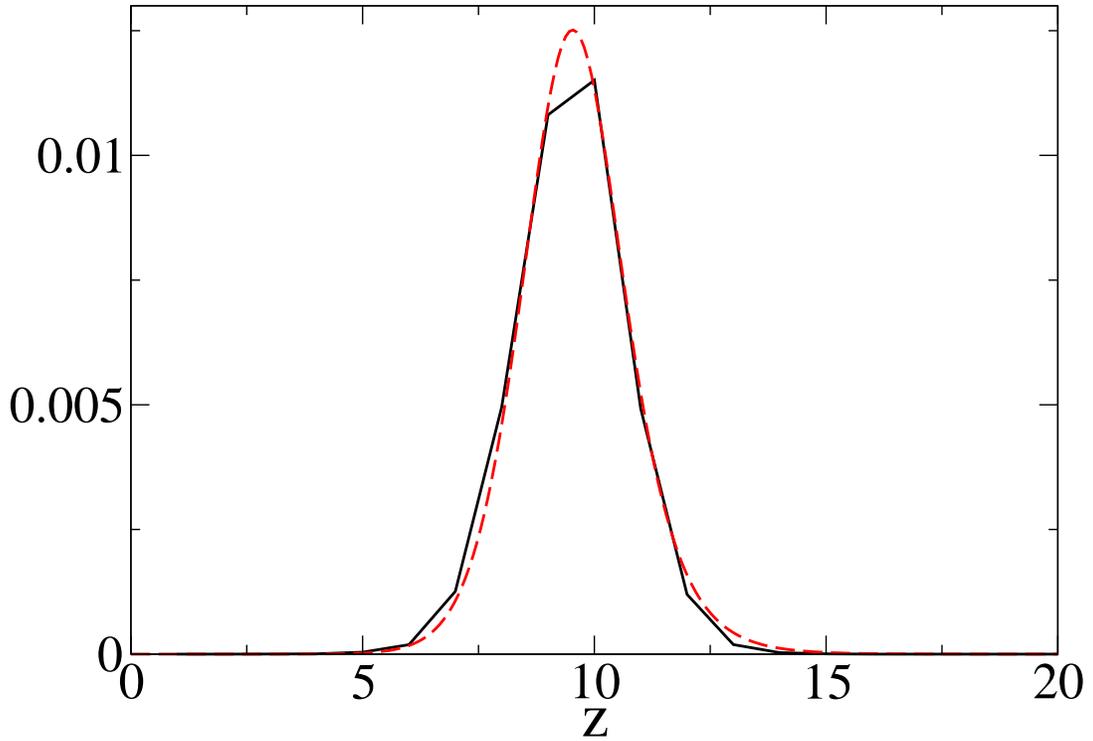}\\
\end{center}
\vspace{-0.5cm}
\caption {The scalar density profile (one-dimensional cut) of a zero
mode representing one of the $82$ heavy dyons (solid line) fitted by
the analytical expression for the scalar density of a dyon's fermion
zero mode (dashed line).}
\label{fig:profile}
\end{figure*}

We evaluated the summed scalar density $\rho(t,{\vec x})$ of all zero and 
nonzero modes. There was a number of clusters of this density exactly equal 
to the number of zero and near-zero modes. First in each cluster the absolute 
maximum was found at ${\vec x}_{\rm max}$:
\be
\rho_{\rm max} = \max_t \max_{\vec x} \rho(t,{\vec x}) \, .
\label{eq:maximum}
\ee
Then in the same spatial point ${\vec x}_{\rm max}$ the minimum with respect 
to $t$ was determined:
\be
\rho_{\rm min} = \min_t \rho(t,{\vec x}_{\rm max}) \, .
\label{eq:minimum}
\ee
From the analytical expression for the zero mode of a one-caloron 
solution~\cite{ferm1} the maximum (\ref{eq:maximum}) and the minimum 
values (\ref{eq:minimum}) of its scalar density can be derived and 
expressed as functions of parameters $\omega$ and $d$. Using these 
functions we determined the parameters from the values of $\rho_{\rm max}$
and $\rho_{\rm min}$ observed for the individual maxima of the scalar
density that are considered as being related to exact zero modes or to
zero modes that are mixed to form the pairs on near-zero modes. In this 
way we have obtained the distribution of the $82$ antiperiodic 
(``heavy'') fermionic modes over their mass fraction $m_{h}$. 
The corresponding histogram is presented in Fig.~\ref{fig:nudistr}. 
Note that for most of the distribution $m_{h} > 0.5$, and that
the maximum is located at about 0.7 which approximately reconfirms 
the value $m_{h} = 0.6$ obtained in section \ref{sec:spectra} under 
the assumption that the holonomy is given by $\langle L \rangle$. 
The typical profile of a fermionic mode, represented 
by its scalar density taken at $t = t_{\rm max}$, $x = x_{\rm max}$, and 
$y = y_{\rm max}$ as a function of $z$ is shown in Fig.~\ref{fig:profile} 
together with the best-fitting profile of the analytically given scalar 
density of a fermion zero mode.

\begin{figure*}[!htb]
\begin{center}
\includegraphics[width=.7\textwidth]{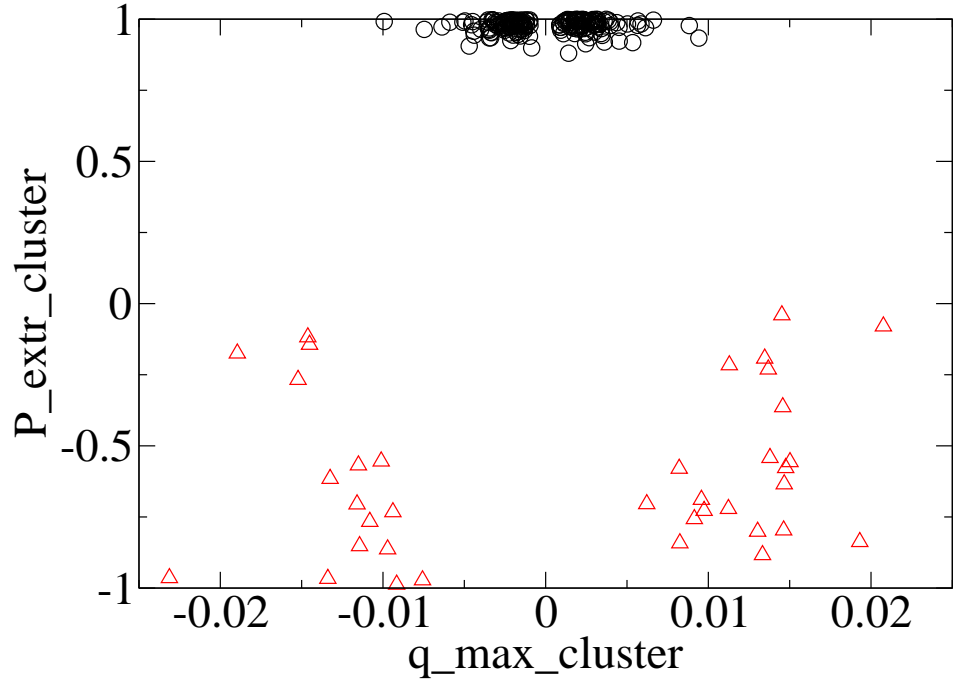}\\
\end{center}
\vspace{-0.5cm}
\caption {Clusters containing static monopoles are shown in a scatter plot 
with respect to the extremal value of the topological charge density and 
the peak value of the local Polyakov line (inside the clusters, including 
the sign). Circles correspond to clusters found by periodic fermions (light
dyons), triangles correspond to clusters found by antiperiodic fermions
(heavy dyons).}
\label{fig:qpcorr}
\end{figure*}

A more complete description of the dyonic structure of the topological objects
found with the help of periodic and antiperiodic fermionic modes can be 
given in terms of gluonic observables. From the analytical solution we know 
that dyons are magnetic monopoles and that the Polyakov loop is peaked with 
a positive sign of $P({\vec x})$ if the dyon supports a periodic fermion 
zero mode and with a negative sign of $P({\vec x})$ if the dyon supports an 
antiperiodic fermion zero mode~\cite{KvB-1,ferm1}. The gluonic configurations
of the present investigation have been subject to smearing as described in 
Ref.~\cite{BIMMMV07}, by 10 steps of APE smearing~\cite{DHK}. We made sure that 
this procedure does not change the low lying spectrum of overlap fermions. After 
smearing the Abelian magnetic monopole content and the Polyakov loop profile 
of the topological clusters have been recorded. Clusters with static Abelian 
monopoles occupy about 3\% of the lattice volume and contain about 50\% of 
all timelike Abelian monopole links. They have peaked values of the Polyakov 
loop correlated in sign with the fermionic boundary condition, in total 
agreement with the required dyon properties. These clusters are recorded in 
Fig.~\ref{fig:qpcorr} in form of a scatter plot with respect to the maximal 
value of the topological charge density and the extremal value of the local 
Polyakov loop $P({\vec x})$ inside the cluster (both including the sign).
One can see that all heavy dyon clusters, denoted by triangles, have 
negative $P({\vec x})$ and relatively large values of the maximal topological 
charge density. The points representing light dyon clusters are 
concentrated with their $P({\vec x})$ close to $+1$ with a maximal topological 
charge density (here in terms of the gluonic topological charge density) 
not exceeding $|q_L(x)| = 0.01$ .

\section{Conclusions}
\label{sec:conclusions}
\vspace{-0.2cm}

We presented in this paper new evidence for the dyonic nature of the 
topological fluctuations, this time for the deconfinement phase of 
$SU(2)$ lattice gauge theory. 
We show that the dyonic picture suggests an explanation of the
strong difference in the spectrum of the overlap Dirac operator with
periodic and antiperiodic boundary conditions, which is the only
explanation given to this phenomena so far. Furthermore, we show
that the abundances of near-zero 
modes, the localization properties of zero modes and near-zero modes, 
the (anti)selfduality properties of topological clusters, the profiles 
of fermionic modes, the monopole content and the Polyakov loop profiles 
of topological clusters are in reasonable agreement with the dyonic picture 
of the topological objects in the ``vacuum'' above $T_c$, where light dyons 
(and light antidyons) are most abundant and heavy dyons (and heavy antidyons) 
are suppressed. All of these topological clusters have properties known from 
asymmetric caloron solution with only slightly nontrivial holonomy. Maximally 
nontrivial holonomy is realized only in the confinement phase, with a full 
symmetry between all types of dyons and antidyons. We have considered in 
particular a temperature $T = 1.5~T_c$ where the average Polyakov loop 
determines the deviation from both limits of trivial and maximally nontrivial 
holonomy. We expect that at higher temperatures the asymmetry between light 
and heavy (anti)dyons in the vacuum will further increase. Asymptotically, 
the topological objects in the vacuum will be exclusively light dyons and 
antidyons appearing in equal number.

\vspace{-0.2cm}
\section*{Acknowledgements}
This work was partly supported by the DFG grant 436 RUS 113/739/0-2 
together with the RFBR-DFG grant 06-02-04010.  
Two of us (V.G.~B. and B.V.~M. ) gratefully appreciate the support of 
Humboldt-University Berlin where this work was carried out to a large 
extent. V.G.~B. is supported by grants RFBR 08-02-00661 and 
RFBR -7-02-00237a. E.-M.~I. was supported by DFG (FOR 465 / Mu932/2). 
He is grateful to the Karl-Franzens-Universit\"at Graz for the guest 
position he holds while the paper is being completed. We thank Falk 
Bruckmann and Christof Gattringer for comments on a draft version of 
this paper.



. 

\end{document}